\documentclass[12pt]{article}
\usepackage{epsf}
\usepackage{amssymb}
\renewcommand{\vector}[1]           % Vector in bold italic
 {\mbox{\boldmath$#1$}}
\newcommand{\rme}{\mathrm{e}}
\newcommand{\rmi}{\mathrm{i}}
\newcommand{\rmd}{\mathrm{d}}
\newcommand{\etal}{{\itshape et~al.}}
\begin{document}
\title{Many-electron tunneling in atoms}
\author{B.~A.~Zon\\
\textit{Voronezh State University}\\
\textit{394693, Russia, Voronezh, University Sq., 1}\\
\textit{E-mail:zon@niif.vsu.ru}}
\date{}
\maketitle
\begin{abstract}
A theoretical derivation is given for the
formula describing $N$-electron ionization of atom by
a dc field and laser radiation in tunneling regime.
Numerical examples are presented for noble gases atoms.
\end{abstract}
{\small
\textbf{PACS numbers:}
32.80.Fb, % Photoionization of atoms and ions
87.64.Lg. % Electron and photoelectron spectroscopy
}

\newpage
\section{Introduction}
Many-electron ionization of atoms by laser field was
firs observed by Suran and Zapesochny~\cite{Sur}
in alkali-earth atoms (the review of that work as
well as some earlier ones see in~\cite{DSZ}).
At present, such studies form one of main guidelines
in physics of strong field interaction
with atoms~\cite{Agos}.

A number of theoretical models were proposed for
interpretation of the gathered experimental data.
Some models dealt with direct influence
of laser radiation on the atomic
electrons~\cite{Kul,Gross,Lam,F,Gol}, the others
consider highly-stripped ion formation due to
nonelastic scattering of previously emitted electrons
with the parent ion~\cite{Rhodes,Kuchiev,Corkum}.
These models allow to explain a number of observed
features of the
phenomenon~\cite{Walker,Chin1,Chin2,Chin3,Chin4}.
Nevertheless, there are some difficulties in
theoretical description of highly-stripped ion
formation in laser field which is not related
to nonelastic collisions~\cite{Kul,Gross,Lam,F,Gol}.
Due to these difficulties, the above mechanisms
cannot be properly used for explanation of
the experiment.

At the same time, it is well-known fact that
the single-charged ion formation by a laser field
in tunnelling regime can be satisfactory described
in terms of relatively simple formulae of the
ADK theory~\cite{SCh,PPT,ADK}.
An empirical generalization of the ADK formulae
for describing the highly-stripped ion formation
was proposed in~\cite{San}. So it would be reasonable
to generalize the available theory of tunnelling
in atoms to the case of non-sequentional multiple
ionization of atom. Solution of this problem is
the objective of the present work.

Obviously, the Josephson effect can be considered
as a solid-state analogue of the considered phenomenon.
Some considerations on difference between the one-
and many-particle tunnelling are mentioned in
reference~\cite{Zah}. Comparison of these considerations
with the results of the present work shows that
the mentioned difference for tunnelling  in atoms is not
so trivial as it was described in~\cite{Zah}.

\section{Asymptotics of the many-electron wave function}

Let us remembers some facts which make the main proposed
concepts easer to understand. To describe optical transitions
in complex atoms, Bates and Damgaard~\cite{BD} modified
the Slater method~\cite{Gombas}. Basically, the
nodeless character of Slater orbitals was retained.
Unlike the Slater method, the effective nuclear charge
ceases to be a fitting parameter for valence electrons
in atom since it coincide with the residual ion charge.
But the effective principal quantum number is uniquely
determined by the electron coupling energy.
So the asymptotical region of electron motion is
considered, where the atomic potential has Coulomb shape.
High accuracy of oscillator strengths calculations~\cite{Sob}
using the Bates--Damgaard method and its clear physical
justification allows to use this method in
calculations of other atomic characteristics determined by
large electron-nucleus distances.

The tunnelling probability is also determined
by large electron-nucleus distances where the
energy of the electron interaction with the external
field becomes comparable with the attractive
energy of the residual ion. So the Bates--Damgaard method
can be used for describing the tunnelling effect. Such a
procedure was developed in recent work~\cite{Z} for
tunnelling calculation in Rydberg molecules. In that
work some evaluations are presented for the
applicability conditions of the method.

Let $N$ equivalent (i.~e. belonging to the same
atomic shell) electrons are removed from the atom via
tunnelling. Then the asymptotic behaviour of the
radial part of $N$-electron wavefunction in the
Bates--Damgaard approximation is determined by the
product of properly symmetrized one-electron
function asymptotics:
%%%%
\begin{eqnarray}
\psi_{\nu lm}(\vector{r})
\sim
C_{\nu l}b^{-3/2}\left(\frac{r}{b}\right)^{\nu-1}
\exp\left(-\frac{r}{b}\right)
Y_{lm}\left(\frac{\vector{r}}{r}\right),
\label{Eq:PsiAsymptotics}%\eqno(1)
\\
C_{\nu l}=(2\pi\nu)^{-1/2}
\left(\frac{2}{\nu}\right)^{\nu}L(\varepsilon),
\qquad
L(\varepsilon)=\left(\frac{1-\varepsilon}{1+\varepsilon}\right)^
{\frac{1}{2}(l+1/2)}(1-\varepsilon^2)^{-\nu/2}.
\nonumber
\end{eqnarray}
%%%%
Here $b=a\nu/Z$, $Z$ is the residual ion charge, $a=\hbar^2/\mu e^2$ is
Bohr radius, $\mu$, $e$ are the mass of electron and the
absolute value of its charge, $\varepsilon=(l+\frac12)/\nu$.
The $C_{\nu l}$~constant in~(\ref{Eq:PsiAsymptotics})
is determined in quasiclassical approximation
not implying the condition~$l\ll\nu$, which was required
in~\cite{ADK}. It results in the arising of
$L(\varepsilon)$ function with $L(\varepsilon)\rightarrow 1$ at
$\varepsilon\rightarrow0$. After passage to this limit the
expression~(\ref{Eq:PsiAsymptotics}) for the
$C_{\nu l}$~constant turns into the formula~(11)
of the reference~\cite{ADK} (with an inaccuracy corrected:
the number $\rme=2.718\dots$ should be omitted).

The expression~(\ref{Eq:PsiAsymptotics}) for $C_{\nu l}$ is
obtained under $\varepsilon<1$. For $\varepsilon>1$,
the quasiclassical approximation is not valid, so
calculation of $C_{\nu l}$ requires numerical
approaches (see, e.~g.,~\cite{RS}).

The principal quantum number~$\nu$ is determined by the electron
coupling energy. Denoting the first, second etc. ionization
potentials of the atom as $E_1/e$, $E_2/e\dots$, the principal
quantum number of $j$-th removed electron is
\begin{displaymath}
\nu_j=\left(\frac{2aE_j}{Z^2e^2}\right)^{-1/2}.
\end{displaymath}
If the electron are equivalent and are
\emph{simultaneously} removed from the atom, then
for all the electrons
%%%\eqno(2)
\begin{equation}\label{Eq:nu}
\nu=\left(\frac{2aE_N}{NZ^2e^2}\right)^{-1/2},
\end{equation}
%%%
where
\begin{displaymath}E_N=\sum_{j=1}^NE_j\end{displaymath}
is the coupling energy of $N$~electrons. Note that in
framework of the considered model, the asymptotic behaviour
of the bound electron wave function~(\ref{Eq:PsiAsymptotics})
\emph{depends} on the number of the removed electrons.
So a partial account is provided for many-electron effects
in the initial state.

Now we consider $N$-electron ionization as removal of
a $N$-electron ``bundle'' -- a peculiar kind of
quasiparticle of mass~$N\mu$ and of charge~$-Ne$.
In the region which determines the ionization process,
we consider the distances between the electrons in the
bundle to be much less than the separation between
the atomic core and the center of bundle mass.
Denoting the distance between the $i$-th and $j$-th electrons
as $\vector{x}_{ij}$, and the position of
the center of bundle mass as $\vector{R}$, we write
the corresponding inequality:
%%%\eqno(3)
\begin{equation}\label{Eq:x<<R}x_{ij}\ll R.\end{equation}
%%%
Since the atom--laser radiation interaction is considered in
dipole approximation, the influence of the field on
$N$~individual electrons is completely equivalent to the influence
of the field on a quasiparticle of charge~$-Ne$ which is located
at the point~$\vector{R}$. As for the interaction of this
quasiparticle with the core Coulomb field, the correspondent
error value is $\sim(x_{ij}/R)^2$, which is small due to
the accepted inequality~(\ref{Eq:x<<R}).

For the mathematical description of the considered model,
one should solve a problem which is analogous to
that is occurred, e.~g. in nuclear $\alpha$-decay theory.
This problem is to construct the quasiparticle wave
function~$\Psi^{(N)}_{\{\nu lm\}}(\vector{R},\{\vector{x}_i\})$
at large distances from the residual system,
using the one-particle wave functions of the system in the
initial state. Symbols in the braces are sets of
quantum numbers or coordinates of individual particles.
To solve this problem we consider the asymptotics of
the function~$\Psi^{(N)}_{\{\nu lm\}}$ at $R\rightarrow\infty$,
which is a product of the one-electron function
asymptotics~(\ref{Eq:PsiAsymptotics}). It is easy to see that
the radial dependencies of the
functions~(\ref{Eq:PsiAsymptotics}) bring the factor
%%\eqno(4)
\begin{displaymath}
\exp\left(-\frac{NR}{b}\right)
\left(\frac{R}{b}\right)^{N(\nu-1)}.
\end{displaymath}
into the asymptotics of~$\Psi^{(N)}_{\{\nu lm\}}$.
To obtain the angular dependence, the mean of the
$\vector{R},\{\vector{x}_i\}$ variables should be detalized.
Since the problem has the axial symmetry for the linearly
polarized field, the orbital moment projections of
non-interacting electrons onto the polarization direction
are conserved. So it is convenient to leave the azimuth angles
$\varphi_i$ the same that in the original spheric coordinate
system centered in the atomic nucleus. The change of variables
will effect only on the absolute values~$\{r_i\}$ and polar
angles~$\{\theta_i\}$. At $\theta\rightarrow 0$, the behaviour
of the Legendre polynomials
involved in the spheric functions~(\ref{Eq:PsiAsymptotics}), is
determined by
\begin{displaymath}
P_l^{|m|}(\cos\theta_i)\sim
(-1)^{|m|}\frac{\sin^{|m|}\theta_i}{2^{|m|}|m|!}
=(-1)^{|m|}\frac{(r_i^2-r_{iz}^2)^{|m|/2}}{2^{|m|}|m|!r_i^{|m|}}
\end{displaymath}
Substituting here $r_i\rightarrow R$, $r_{iz}\rightarrow R_z$
and introducing the parabolic coordinates $\xi=R+R_z$,
$\eta=R-R_z$ for the center of the bundle mass, the asymptotics
of the $N$-electron function at $\xi\gg\eta$ can be written in
the form
%%%%
\begin{eqnarray}
\Psi^{(N)}_{\{\nu lm\}}(\vector{R},\{\vector{x}_i\})=
B\phi(\xi,\eta)\chi(\{r_i,\theta_i\})
\prod_{j=1}^N\frac{1}{\sqrt{2\pi}}\exp(\rmi m_j\varphi_j),
\nonumber\\
B=a^{-3/2}C_{\nu l}^N\left(\frac Z{\nu}\right)^{3N/2}(2l+1)^N
\prod_{j=1}^N\frac{(-1)^{|m_j|}}{|m_j|!}\left[\frac{(l+|m_j|)!}{(l-|m_j|)!}
\right]^{1/2},
\label{Eq:B}%\eqno(5)
\\
\phi(\xi,\eta)\sim\exp\left[-\frac{N(\xi+\eta)}{2b}\right]
\left(\frac{\xi}{2b}\right)^{N(\nu-1)}
\left(\frac{\eta}{\xi}\right)^{M/2},
\qquad
M=\sum_{j=1}^N|m_j|.
\nonumber
\end{eqnarray}
%%%%
Here $\chi$ is the normalized per unit wave function
of the electron inner motion
in the bundle. Note that there are only $2(N-1)$
independent variables~$\{r_i,\theta_i\}$ of $2N$.
The function~$\phi(\xi,\eta)$ describes the motion
of the center of the bundle mass.

\section{Tunnelling probability}

The further calculation of the tunnelling probability
is implemented according the standard technique~\cite{SCh,LL},
an account provided for that the electron bundle mass
is~$N\mu$ and its charge is~$-Ne$. Substituting the
function~$\phi(\xi,\eta)$ from (\ref{Eq:B}) into the Schr\"odinger equation
%%\eqno(6)
\begin{displaymath}
\frac{\rmd}{\rmd\xi}\left(\xi\frac{\rmd\phi}{\rmd\xi}\right)+
\left(\beta-\frac{E_NN\mu}{2\hbar^2}\xi\right)\phi=0,
\end{displaymath}
describing the motion with respect to the parabolic~$\xi$
coordinate at $\xi\rightarrow\infty$,
we obtain the variables separation constant:
%%%\eqno(7)
\begin{equation}\label{Eq:beta}
\beta=\frac{N}{b}\left[N(\nu-1)-\frac{M-1}{2}\right].
\end{equation}
%%%
The centrifugal potential is neglected since it vanishes
rapidly at $\xi\rightarrow\infty$.

Now we consider the external field~$F(t)$ to be slow-varying,
and use quasiclassical approximation for the wave
function~$\phi_F(\xi,\eta)$ which describes the center of the
bundle mass motion in the field.
In the below-threshold domain
%%%%
\begin{eqnarray}
\phi_F(\xi,\eta)=
\varkappa(\xi|p(\xi)|/\hbar)^{-1/2}\exp\left(\frac{1}{\hbar}
\int_{\xi_1}^{\xi}|p(\xi)|\rmd\xi\right),
\label{Eq:phi_F}%\eqno(8)
\\
p(\xi)=\hbar\left(-\frac{E_NN\mu}{2\hbar^2}+\frac{\beta}{\xi}+
\frac{1}{4\xi^2}+\frac{N^2e\mu}{4\hbar^2}F\xi\right)^{1/2},
\end{eqnarray}
%%%%
where $\xi_1$ is the greater root of the equation~$p(\xi)=0$.
Comparing the expression~(\ref{Eq:phi_F}) with the function
$\phi(\xi,\eta)$ from~(\ref{Eq:B}) at the point~$\xi_0$
lying in the region
%%%\eqno(9)
\begin{equation}\label{Eq:xi_interval}
\frac{2\hbar^2\beta}{E_NN\mu}\simeq b
\ll\xi_0\ll
\frac{2E_N}{NeF}=\frac{eZ}{b\nu F},
\end{equation}
%%%
we obtain the $\varkappa$ value:
%%%\eqno(10)
\begin{equation}\label{Eq:kappa}
\varkappa(\eta;\xi_0)\simeq
\left(
\frac{N\xi_0}{2b}\right)^{1/2}\exp\left(-\frac{1}{\hbar}
\int_{\xi_0}^{\xi_1}|p(\xi)|\rmd\xi
\right)
\phi(\xi_0,\eta).
\end{equation}
%%%
The condition of existence of the region~(\ref{Eq:xi_interval})
leads to following restriction to the external field:
%%%\eqno(11)
\begin{equation}\label{Eq:F<<Fa}
F\ll F_a\equiv\frac{eZ}{b^2\nu}=
\frac{e}{a^2}\left(\frac{Z}{\nu}\right)^3,
\end{equation}
%%%
which differs from the condition arising in the
one-electron tunnelling description only by the definition
of the $\nu$ value. It should be noted that for
$\nu$ essentially greater than 1 (what holds, e.~g. for
Rydberg states) the inequality~(\ref{Eq:F<<Fa}) is changed
by a stronger one:
%%%\eqno(12)
\begin{equation}\label{Eq:F<Z^3e/16nu^4a^2}
F<\frac{Z^3e}{16\nu^4a^2},
\end{equation}
%%%
which is deduced from the condition of existence of the
potential barrier~\cite{B}.

\begin{figure}[!h]
\epsfbox{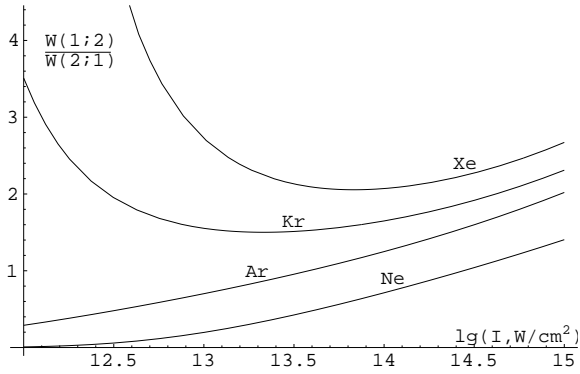}
\caption{
Relation of the 3-charged ion formation probabilities
for noble gases atoms by two different channels
(see~the text).}
\label{Fig:1}
%\vspace{35 ex}
\end{figure}

The formulae~(\ref{Eq:phi_F}) and~(\ref{Eq:kappa}) determine
the function~$\phi_F(\xi,\eta)$ outside the barrier.
With the account of inequality~(\ref{Eq:xi_interval}),
its squared absolute value is~\cite{LL}
%%%
\begin{eqnarray}
|\phi_F(\xi,\eta)|^2
&=&
\frac{\hbar N\xi_0}{2b\xi p(\xi)}
\left(\frac{\xi_0}{2b}\right)^{2N(\nu-1)}
\left(\frac{\eta}{\xi_0}\right)^M
\label{Eq:|phi_F|^2}%\eqno(13)
\\
&\times&
\exp\left[-\frac{N\eta}{b}-\frac{16\hbar^2}{3N^2\mu eF}
\left(\frac{E_NN\mu}{2\hbar^2}\right)^{3/2}-
\beta\left(\frac{2\hbar^2}{E_NN\mu}\right)^{1/2}
\log\frac{NeF\xi_0}{8E_N}\right].
\nonumber
\end{eqnarray}
%%%%
Using~(\ref{Eq:nu}) and~(\ref{Eq:beta}), it is easy
to see that the dependence on the arbitrary parameter $\xi_0$
is actually disappeared in~(\ref{Eq:|phi_F|^2}):
%%%\eqno(14)
\begin{equation}\label{Eq:|phi_F|^2_without_xi0}
|\phi_F(\xi,\eta)|^2=\frac{\hbar N(\eta/b)^M}{2^M\xi p(\xi)}
\left(\frac{2F_a}{F}\right)^{2N(\nu-1)-M+1}
\exp\left(-\frac{N\eta}{b}-\frac{2NF_a}{3F}\right).
\end{equation}
%%%
The ionization probability is determined by the flux
of probability density~(\ref{Eq:|phi_F|^2_without_xi0})
through a plane perpendicular to $z$-axis~\cite{LL}:
\begin{displaymath}
W_{\nu l}^{(N)}(F)\sim
2\pi\int_0^{\infty}v_z|\phi_F(\xi,\eta)|^2\rho\,
\rmd\rho,
\qquad
v_z=\frac{2p(\xi)}{N\mu},
\quad
\rho=\sqrt{\xi\eta},
\quad\rmd\rho\simeq\sqrt{\frac{\xi}{\eta}}\rmd\eta.
\end{displaymath}
Substituting here the formulae~(\ref{Eq:B})
and~(\ref{Eq:|phi_F|^2_without_xi0}), we obtain:
%%%%
\begin{eqnarray}
W_{\nu l}^{(N)}(F)
&=&
\frac{\pi\hbar}{a^2\mu}
\frac{M!(2l+1)^NC_{\nu l}^{2N}}{2^{M-2}N^{M+1}}
\left(\frac{Z}{\nu}\right)^{3N-1}
\prod_{j=1}^N\frac{(l+|m_j|)!}{(|m_j|!)^2(l-|m_j|)!}
\nonumber\\
&\times&\left(\frac{2F_a}{F}\right)^{2N(\nu-1)-M+1}
\exp\left(-\frac{2NF_a}{3F}\right).
\label{Eq:Wnul}%\eqno(15)
\end{eqnarray}
%%%%
This formula determines the $N$-electron tunnelling
probability in dc field within a factor accounting for
the overlapping of wave functions of the electrons remaining
in the atom, with the wavefunctions of the same electrons in the
initial state. Obviously, this factor cannot exceed 1, and its more
accurate evaluation can be performed only numerically.
Note that the $N$ multiplier in the exponent in~(\ref{Eq:Wnul})
in now ways gives an exhaustive account for the dependence
of this exponent on~$N$, as it was considered in~\cite{Zah}.
Due to the formulae~(\ref{Eq:nu}),~(\ref{Eq:F<<Fa}),
this dependence is significantly more complicated and it
is determined by the spectrum of the particular atom.
We present below (figure~\ref{Fig:1}) a numerical example
illustrating this statement.

%%%%%
%%%%%

Now we consider that
%%%%\eqno(16)
\begin{equation}\label{Eq:F}F(t)=F_0\cos\omega t,\end{equation}
where $\omega$ is the laser field frequency. It is a well-known
fact that the tunnelling in a laser field is possible
for small values of the Keldysh parameter~\cite{Keld}
\begin{displaymath}
\gamma=\frac{\sqrt{2\mu E_1}}{eF}\omega,
\end{displaymath}
where $E_1$ is the coupling energy of one electron. Following
the technique developed in~\cite{Keld} for ``particle'' of
mass $N\mu$ and charge $-Ne$, it is easy to see that
the $N$-electron tunnelling is possible for small values
of the parameter
%%\eqno(17)
\begin{equation}\label{Keld}
\gamma_N=\frac{\sqrt{2\mu E_N/N}}{eF}\omega.
\end{equation}
Since the coupling energy is increasing for each
subsequent electron, $N$-electron tunnelling requires
field values lower than $N$-electron tunnelling
cascade.

Substituting~(\ref{Eq:F}) into~(\ref{Eq:Wnul}), we average the
result over the time interval $t\in
[-\pi/2\omega,\pi/2\omega]$~\cite{PPT}%
\footnote{The values
$t\in [\pi/2\omega,3\pi/2\omega]$ leads to $F(t)<0$ and the
tunnelling takes place in the direction of negative $z$
semiaxis.}.
Due to the inequality~(\ref{Eq:F<<Fa}), the
integral arising here can be calculated using the saddle-point
method.  Under the condition~(\ref{Eq:F<Z^3e/16nu^4a^2})
fulfilled, the saddle point is $t=0$, and the final formula is:
%%%%
\begin{eqnarray}
W_{\nu l}^{(N)}(F_0) &=&
\frac{\sqrt{3\pi}\hbar}{a^2\mu}\frac{M!(2l+1)^N C_{\nu
l}^{2N}}{2^{M-3/2}N^{M+3/2}}\left(\frac Z{\nu}\right)^{3N-1}
\prod_{j=1}^N\frac{(l+|m_j|)!}{(|m_j|!)^2(l-|m_j|)!}
\nonumber\\ &\times&
\left(\frac{2F_a}{F_0}\right)^{2N(\nu-1)-M+1/2}
\exp\left(-\frac{2NF_a}{3F_0}\right).
\label{Eq:W(N)}%\eqno(18)
\end{eqnarray}
%%%%
Remember that the exponent dependence on $N$ in~(\ref{Eq:W(N)})
is not reduced to the factor~$N$ which is written explicitly.

\section{Numerical examples}

Unfortunately, the obtained formulae cannot be
immediately related to an experiment, because,
along with the direct $N$-fold ions formation, there
are a number of cascade processes as well as other
ionization mechanisms due to nonelastic collisions of
electrons and ions~\cite{Rhodes,Kuchiev,Corkum}.
For the relation of the theory with an experiment,
the correspondent kinetic equations are to be solved,
that should be a subject for another work.
So only some illustrative examples are considered in this
section.

The figure~\ref{Fig:1} presents the relation of
probabilities of 3-fold ions formation in
the noble gases resulted from two 2-cascade processes:
$\mathrm{A}
\rightarrow\mathrm{A}^+\rightarrow\mathrm{A}^{3+}$ and
$\mathrm{A}
\rightarrow\mathrm{A}^{2+}\rightarrow\mathrm{A}^{3+}$.
These probabilities are denoted as $W(1;2)$ and $W(2;1)$
correspondingly. They have similar dependence on the
laser pulse duration. As it is seen, the relation
$W(1;2)/W(2;1)$ is not equal to 1, as it is follows
from the results of reference~\cite{Zah}.

The following result seems to be curious. The 2-electron
tunnelling probabilities for neutral atoms can be greater than
the one-electron tunnelling probabilities in correspondent
singly charged ions. E.~g., for Ar atom the  2-electron
tunnelling probability exceeds the 1-electron process
probability for $\mathrm{Ar}^+$ ion at the
intensities~$I>10^{14.88}\,\mathrm{W/cm}^2$.  The same result
takes place for~Kr at $I>10^{14.76}\,\mathrm{W/cm}^2$, for~Xe
at $I>10^{14.34}\,\mathrm{W/cm}^2$.  At the same time, for
light noble gases atoms~He and~Ne, the probabilities of
one-electron tunnelling in singly charged ions are approximately
by two orders greater than the probabilities of two-electron
process in the correspondent neutral atoms at~$I\simeq
10^{15}\,\mathrm{W/cm}^2$. These facts shows wide range of
experimental situations arising in multiphoton tunnelling
effect.

This work was stimulated by the report~\cite{San}. The author
is grateful to Professor W.~Sandner for the interest to the
work, and to WE--Heraeus-Stiftung for the offered opportunity
to participate in the seminar work. I also express my deep
gratitude to Professor N.~B.~Delone and to the participants of
his seminar in IOF~RAN for helpful discussion.
This work was partially supported by Russian Foundation for
Basic Researches (grant no.~97-02-18035).

\end{document}